\documentclass[aps,prl,twocolumn,superscriptaddress,showpacs,preprintnumbers,amsmath,amssymb]{revtex4-1}
\usepackage{graphicx}
\usepackage{dcolumn}
\usepackage{bm}
\usepackage{times}

\newcommand\be{\begin{equation}}
\newcommand\ee{\end{equation}}

\begin{document}
\title{
Elongated Fermi superfluid:
absence of critical imbalance enhancement at equilibrium}
\author{Masaki Tezuka}
\email{tezuka@scphys.kyoto-u.ac.jp}
\affiliation{Department of Physics, Kyoto University,
Kitashirakawa-oiwakecho, Sakyo-ku, Kyoto 606-8502, Japan}
\author{Youichi Yanase}
\email{yanase@phys.sc.niigata-u.ac.jp}
\affiliation{Department of Physics, Niigata University,
8050 Ikarashi, Nishi-ku, Niigata 950-2181, Japan}
\author{Masahito Ueda}
\email{ueda@phys.s.u-tokyo.ac.jp}
\affiliation{Department of Physics, University of Tokyo,
7-3-1 Hongo, Bunkyo-ku, Tokyo 113-0033, Japan}
\affiliation{ERATO Macroscopic Quantum Control Project, JST, Tokyo 113-8656, Japan}
\date{\today}

\begin{abstract}
We show that the maximum population imbalance ratio $P_\mathrm{CC}$
for a two-component Fermi gas near the unitarity limit to condense
does not increase with the trap aspect ratio $\lambda$,
by two methods of
1) solving the Bogoliubov-de Gennes equations with coupling-constant renormalization, and
2) studying the pairing susceptibility by the real-space self-consistent $T$-matrix approximation.
The deviation of the cloud shape from what is expected from the trap shape increases
but stays minor with increasing $\lambda$ up to $50$.
This finding indicates that despite the apparent discrepancy between the MIT and Rice
experiments over the value of $P_\mathrm{CC}$ and the validity of local density
approximation, the equilibrium state of the system for the aspect ratio in the
Rice experiment should be consistent with that of MIT.
\end{abstract}

\pacs{03.75.Ss, 71.10.Ca, 37.10.Gh}
\maketitle

Gaseous Fermi superfluids are endowed with new degrees of controllability over population difference and trap anisotropy.
Imbalanced superfluidity of $^6\mathrm{Li}$ has been observed by the Rice \cite{RiceExpl} and MIT \cite{MITExpl} groups, but their results have shown marked differences over the validity of local density approximation (LDA) and the Chandrasekhar-Clogston (CC) limit -- the upper bound
$P_\mathrm{CC}$ of imbalance parameter $P\equiv(N_\uparrow-N_\downarrow)/N$ beyond which superfluidity breaks down \cite{CClimit}, where $N_\uparrow$ and $N_\downarrow$ are the numbers of majority and minority atoms, and $N\equiv N_\uparrow+N_\downarrow$ is the total atom number.
In the MIT experiment the profiles of both majority and minority clouds obey LDA, while in the Rice experiment with a very elongated trap and fewer atoms, LDA apparently breaks down.
The CC limit was observed at MIT but not at Rice.
A phenomenological surface tension \cite{DeSilva06, Haque07} of the condensate was shown to reproduce the deformation observed by the Rice group, but how to reconcile the apparently contradicting experimental differences without free parameters remains elusive
\cite{2009PhRvA..79f3628B, Diederix09}.
More recently the non-equilibrium state during the evaporative cooling process
\cite{Parish09} was discussed to explain the Rice results.
In this Letter we demonstrate, for the equilibrium state of the system at low and
finite temperatures, that
1) the CC limit does not increase with increasing the trap aspect ratio $\lambda$,
and that
2) while the density-difference distribution does deform from what is expected from
the trap shape, the deformation is not as significant as in the Rice experiment
for the number of atoms as small as $3\times 10^4$.

We consider a system of atoms with mass $m$ confined in an axisymmetric harmonic potential 
$V(\mathbf{r}) \equiv m(\omega_\perp^2 (x^2+y^2) + \omega_z^2 z^2)/2$
with axial frequency $\omega_z$ and radial one $\omega_\perp$, and analyze superfluidity of this system using the Bogoliubov-de Gennes (BdG) equations \cite{Mizushima05,Castorina05, Kinnunen06,
Machida-Mizushima, PhysRevA.76.033620, Liu07, Sensarma07, Baksmaty10}. 
Sensarma \textit{et al.} \cite{Sensarma07} studied the shape of the atom cloud by changing $N$ and $P(\leq 0.4)$, and argued that $(N/\lambda)^{1/3}\gg1$ ($\lambda\equiv\omega_\perp/\omega_z$) should be the condition for the validity of LDA.
For $(N/\lambda)^{1/3}\sim 10$, the cloud shape obtained
in Ref.~\cite{Sensarma07} looks quite similar to that of the equipotential surface.
However, the Rice experiment shows the breakdown of LDA for almost the same value of $(N/\lambda)^{1/3}$.
While our numerical results also show deformation similar to that found in Ref.~\cite{Sensarma07} for $N\sim10^3$
at $\lambda=4$, the density profiles are different presumably because we incorporate the effect 
of the chemical potential difference as well as the interaction between atoms in the normal
state. Such deformation almost disappears for $N\sim3\times 10^4$.

\begin{figure}
\includegraphics{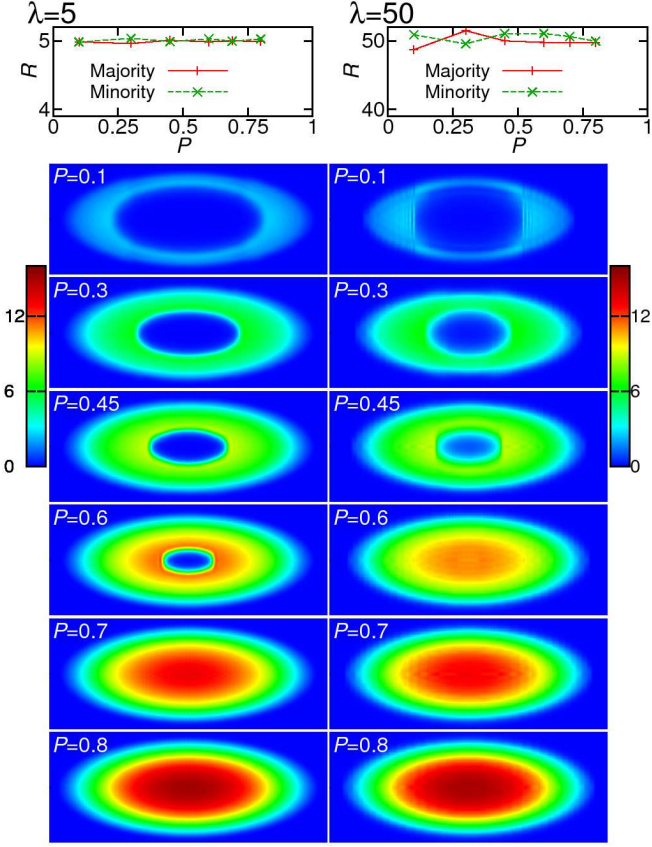}
\caption{(Color online) Top panel shows the $P$ dependence
of the ratio $R$ of the axial to radial cloud widths at which the atomic density equals $1/20$ of its peak value. 
Lower panels show the density difference between majority and minority atoms plotted for $\lambda=5$ (left)
and $50$ (right) with $k_{\rm F}a_s=-1.3$, $N=3\times10^4$, and varying imbalance parameter $P$.
The trap axis lies in the horizontal direction.
At $\lambda=5$, $(\mu_\uparrow,\mu_\downarrow)=(38.83, 28.80)$ for $P=0.1$ and
$(53.08, 4.79)$ for $P=0.8$;
at $\lambda=50$, $(\mu_\uparrow,\mu_\downarrow)=(38.63, 28.78)$ for $P=0.1$ and
$(53.13, 4.66)$ for $P=0.8$.}
\label{fig:r002r02}
\end{figure}

The BdG equations for unequal chemical potentials
$(\mu_{\uparrow}, \mu_{\downarrow})$ are given by
\begin{equation}
\left(
\begin{array}{cc}
\hat H_{\uparrow} + W_{\downarrow} & \Delta\\
\Delta^* & -\hat H_{\downarrow} - W_{\uparrow}
\end{array}
\right)
\left(\begin{array}{c}u_{q}\\v_{q}\end{array}\right)
=
\epsilon_{q}\left(\begin{array}{c}u_{q}\\v_{q}\end{array}\right),
\label{eqn:BdG}
\end{equation}
where
$\hat H_{\sigma} \equiv  -\bm \nabla^2/(2m) + V(\bm r) - \mu_{\sigma}$ 
($\sigma=\uparrow,\downarrow$)
is the one-body Hamiltonian, and $W_\sigma(\bm r)$ is the Hartree-Fock mean-field energy $gn_\sigma(\bm r)$
with the coupling constant $g$ given in terms of s-wave scattering length $a_s$ as $g=4\pi\hbar^2a_s/m$.
In the following we take $m=\hbar=k_\mathrm{B}=1$, set
$\overline{\omega}\equiv\sqrt[3]{\omega_\perp^2\omega_z}=\omega_\perp/\sqrt[3]{\lambda}$,
and choose $\sqrt{\hbar/(m\overline{\omega})}=1$ as the unit of length.
The self-consistent conditions give the density distributions $n_\sigma(\bm r)$ and
the s-wave singlet pair amplitude $\Delta(\bm r)$ as
\begin{eqnarray}
n_{\uparrow}(\bm r) &=& \sum_{q} f_{q} |u_{q}(\bm r)|^2,
n_{\downarrow}(\bm r) = \sum_{q} \left(1-f_{q}\right)
|v_{q}(\bm r)|^2,\nonumber\\
\Delta(\bm r) &=& g_{\rm eff}(\bm r)\sum_{q}f_{q} 
u_{q}(\bm r) v_{q}^*(\bm r),\label{eqn:BdGDensity}
\end{eqnarray}
where $f_q\equiv (e^{\beta\epsilon_{q}}+1)^{-1}$ is the Fermi distribution
function with $\beta\equiv (k_\mathrm{B}T)^{-1}$.
To cope with the ultraviolet divergence in $\Delta(\bm r)$,
we follow Bulgac and Yu \cite{Bulgac02} and treat the contribution from
states above an energy cutoff $E_c$ within LDA.
In Ref. \cite{Bulgac02}, where $\mu=\mu_\uparrow=\mu_\downarrow$ is assumed,
the single-particle Green's function $G_\mu^0$ with
$\hat H_0= -\bm \nabla^2/(2m) + V - \mu$ is used to remove the divergence.
The regular part $G_\mu^{0,\mathrm{reg}}$ of $G_\mu^0$ is obtained by
employing the Thomas-Fermi approximation for the states above $E_c$, so that 
the effective coupling constant is given in terms of 
$k_\mathrm{c}(\bm r) \equiv \sqrt{2\left(E_c-V(\bm r)\right)}$ and
$k_{\rm F}^0(\bm r) \equiv \sqrt{2\left(\mu-V(\bm r)\right)}$ as
\begin{equation}
\frac{1}{g_{\rm eff}(\bm r)}
= \frac{1}{g}+\frac{1}{2\pi^2}\left(\frac{k_{\rm F}^0(\bm r)}{2}
\ln\frac{k_\mathrm{c}(\bm r) + k_{\rm F}^0(\bm r)}{k_\mathrm{c}(\bm r) - k_{\rm F}^0(\bm r)}
- k_\mathrm{c}(\bm r)\right).
\label{eqn:balanceGEff}
\end{equation}
Grasso and Urban \cite{Grasso03} replaced $k_{\rm F}^0(\bm r)$ with
$\tilde k_{\rm F}(\bm r) \equiv \sqrt{2\left(\mu-V(\bm r)-W(\bm r)\right)}$, where
$W(\bm r)=W_{\uparrow,\downarrow}$ for $\mu_\uparrow=\mu_\downarrow$,
so that the convergence is achieved for much smaller values of $E_c$.
We adopt this method except that we replace $G_\mu^{0,\mathrm{reg}}$ by
$\left(G_{\mu_\uparrow}^{0,\mathrm{reg}}+G_{\mu_\downarrow}^{0,\mathrm{reg}}\right)/2$ to 
maintain a given chemical potential difference.
Consequently, Eq. (\ref{eqn:balanceGEff}) is replaced by
\begin{equation}
\frac{1}{g_{\rm eff}(\bm r)}
= \frac{1}{g}+\frac{1}{2\pi^2}
\left(
\sum_\sigma
\frac{\tilde k_{{\rm F}\sigma}}{4}
\ln
\frac{k_\mathrm{c} + \tilde k_{{\rm F}\sigma}}
{k_\mathrm{c} - \tilde k_{{\rm F}\sigma}}
- k_\mathrm{c}\right),
\label{eqn:imbalanceGEff}
\end{equation}
where
$\tilde k_{{\rm F}\sigma}(\bm r)\equiv
\sqrt{2\left(\mu_\sigma-V(\bm r)-W_{\overline{\sigma}}(\bm r)\right)}$,
with $W_\sigma(\bm r) = g n_\sigma(\bm r)$.
While BdG theory was originally proposed to describe the weak-coupling BCS limit, it was demonstrated to describe the BEC limit \cite{Pieri03}, and 
the BCS-BEC crossover region was also studied by this theory \cite{Heiselberg04}. We therefore expect that this theory is applicable, at least qualitatively,
for the strongly interacting region with population imbalance,
provided that an appropriate coupling-constant renormalization is employed.

At the unitarity limit $(k_{\rm F}a_s)^{-1}\rightarrow0$,
the normal state interaction does not diverge, and
the binding energy of a single
$\downarrow$ atom to the Fermi sea of $\uparrow$ atoms
with the Fermi energy $E_{{\rm F}\uparrow}$ is $-(3/5)AE_{{\rm F}\uparrow}$ with $A=0.97(2)$ \cite{LoboPRL2006}.
This corresponds to the mean-field energy of $-(9A\pi/20)\left(k_{{\rm F}\uparrow}(\bm r)\right)^{-1}\times4\pi n_\uparrow(\bm r)n_\downarrow(\bm r)$,
where $k_{\mathrm{F}\sigma} \equiv (6\pi^2 n_\sigma)^{1/3}$.
On the BCS side of the unitarity limit, the normal state interaction should be weaker than at the unitarity limit
so $|k_{{\rm F}\uparrow}(\bm r)a_s| \leq |k_{\rm F}(\bm 0)a_s| < 9A\pi/20 = 1.37$.
Moreover, we can show that the BdG equations (\ref{eqn:BdG}) do not have a stable self-consistent solution
for $|k_{\rm F}a_s|>3\pi/4 = 2.36$ for the homogeneous case without chemical potential difference.

We take $k_{\rm B}T=0.05\hbar\overline{\omega}$ and use the Steffensen iteration to solve Eqs. (\ref{eqn:BdG}) and (\ref{eqn:BdGDensity}),
to self-consistently determine $n_\uparrow(\bm r)$, $n_\downarrow(\bm r)$ and $\Delta(\bm r)$
for a given set of $(\mu_\uparrow, \mu_\downarrow)$.
The number of atoms in the $\sigma(=\uparrow,\downarrow)$ state is defined as
$N_\sigma\equiv\int\mathrm{d}^3rn_\sigma$.

Figure~\ref{fig:r002r02} shows the main results of this Letter.
For both $\lambda=5$ and $\lambda=50$, the ratio of the axial to radial cloud widths remains
close to $\lambda$ for both minority and majority atoms,
and the dip of the density difference rapidly dwindles with increasing $P$,
vanishing for $P>0.75$.
Thus the CC limit is not enhanced as $\lambda$ is increased.
For $\lambda=50$, the density difference shows some deformation for small $P$,
but it disappears for $P\gtrsim0.60$.

\begin{figure*}
\includegraphics{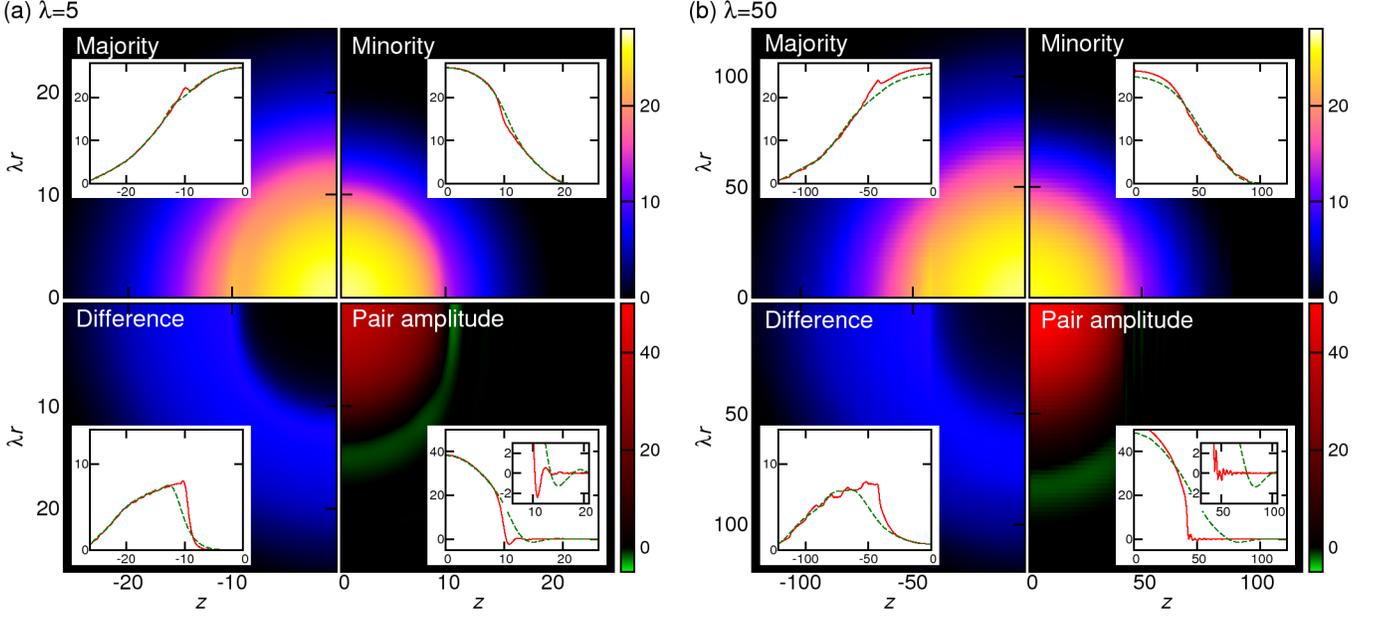}
\caption{
(Color online) 
Majority and minority density distributions $n_{\uparrow,\downarrow}(z,r)$,
their difference $n_\uparrow(z,r)-n_\downarrow(z,r)$
and pair amplitude $\Delta(z,r)$ plotted for (a) $\lambda=5$ and
(b) $\lambda=50$ with $k_{\mathrm{F}\uparrow}a_s=-1.3$ and $N=3\times10^4$ at $P=0.40$.
$(\mu_\uparrow, \mu_\downarrow)=(45.90, 19.20)$ in (a) and $(45.57, 19.42)$ in (b).
The density distributions and $\Delta$ are displayed in color-coded gauges shown on the upper right and lower right, respectively.
In each inset, the cross sections at $r=0$ (solid curve) and $z=0$ (dashed curve) are
plotted against $z$ and $\lambda r$, respectively.
For the pair amplitude, the regions close to the horizontal axis are
enlarged in the smaller insets.
}
\label{fig:Distribution}
\end{figure*}

Figure~\ref{fig:Distribution} shows typical distributions of $n_{\uparrow, \downarrow}$, their difference, and $\Delta$ for $P=0.40$.
We rescale the calculated distribution as $r\rightarrow \lambda r$ so that the equipotential surface becomes a circle.
For $\lambda=5$, the shape of the minority component and the density difference
closely follow the equipotential surface, as shown in the left column of Fig.~\ref{fig:r002r02}.
The pair amplitude shows sign changes,
which are absent in LDA but shows up in the BdG simulation
as discussed in Ref.~\cite{PhysRevA.76.033620} for a spherical system.

For $P$ larger than $0.7$, the pair amplitude almost vanishes, and
the density difference peaks at ${\bm r}={\bm 0}$.
We therefore conclude that LDA is essentially valid at $\lambda=5$
as observed by the MIT group.
For $\lambda=50$, while the density difference shows some deviation from the trap shape,
implying the breakdown of LDA, the degree of breakdown is rather small.
This can be seen from almost spherical density distributions of both the majority and
minority components in Fig.~\ref{fig:Distribution}.
(Note that in Fig.~\ref{fig:Distribution} the vertical axis is scaled by a factor of $\lambda$.)
The region with non-vanishing pairing amplitude is also rather similar to that of the minority component,
reflecting the fact that pairing occurs effectively in the strongly interacting regime.

\begin{figure}
\begin{center}
\includegraphics{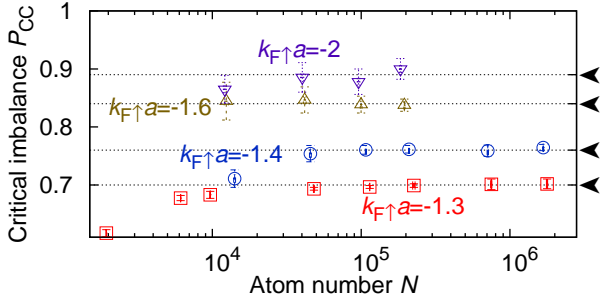}
\end{center}
\caption{
(Color online)
CC limit $P_\mathrm{CC}$ plotted against the total number of trapped atoms
$N$ in a spherical harmonic potential.
Here, $P_\mathrm{CC}$ is identified as the value of $P$ at which the extrapolated
plot of $\Delta(\bm 0)$ crosses zero.}
\label{fig:CClimit}
\end{figure}

With the same number of atoms,
we have thus confirmed that LDA is less invalid at $\lambda=5$ than at $\lambda=50$.
The breakdown of LDA is a finite-size effect, and it is enhanced for larger $\lambda$.
Figure \ref{fig:CClimit} shows the atom-number dependence of
$P_\mathrm{CC}$ for a spherical trap.
We find that with increasing $N$, $P_\mathrm{CC}$ approaches a constant
value for each $|k_{\mathrm{F}\uparrow}a_s|$,
which is, for $|k_{\mathrm{F}\uparrow}a_s|=1.3$, close to the value at which the pair amplitude
disappears in the elongated traps with $\lambda=5$ and $50$.

			
\begin{figure}
\begin{center}
\includegraphics{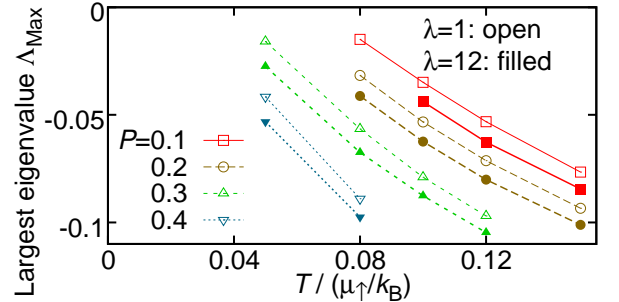}
\end{center}
\caption{
(Color online)
Largest eigenvalue $\Lambda_\mathrm{Max}$ of $\chi_\mathrm{SC}$ obtained in the RSTA method
is plotted against $T/(\mu_\uparrow/k_\mathrm{B})$ for
$\mu_\uparrow=10\hbar\overline{\omega}=10\hbar\sqrt[3]{\omega_z\omega_r^2}$
for aspect ratios $\lambda=1$ (open symbols) and $12$ (filled)
and imbalance parameters $P=0.1, 0.2, 0.3, 0.4$.
}
\label{fig:RSTA}
\end{figure}

To show that the non-increasing behavior of $P_\mathrm{CC}$ for increasing $\lambda$ is not
an artifact of the
BdG approximation or a finite $k_{\mathrm{F}\uparrow} a_s$, 
we employ the real-space self-consistent $T$-matrix approximation
\cite{Yanase06} (RSTA), by which strongly interacting fermions in
an inhomogeneous potential can be treated with high accuracy.
RSTA has been shown to reproduce the pseudo-gap phase in high-$T_\mathrm{c}$ superconductors \cite{Yanase06}
and the superconductor-insulator transition in disordered diamond superconductors \cite{Yanase09}.

At the unitarity limit $(1/(k_{\mathrm{F}}a_s)\rightarrow0)$,
in the normal phase we self-consistently solve the following set of equations:
\begin{eqnarray}
\chi_\mathrm{SC}(\bm r,\bm r')
&=& T\sum_nG_\uparrow(\bm r,\bm r',\omega_n)
G_\downarrow(\bm r,\bm r',-\omega_n)\nonumber\\
&-&C(\bm r)\delta(\bm r,\bm r'),\\
\tilde T(\bm r,\bm r') &=& \left[g^{-1}\delta(\bm r,\bm r')
+ \chi_\mathrm{SC}(\bm r,\bm r')\right]^{-1},\\
\Sigma_\sigma(\bm r,\bm r',\omega_n)
&=&
TG_{\overline{\sigma}}(\bm r',\bm r,-\omega_n)
\tilde T(\bm r,\bm r'),\\
G_\sigma(\bm r,\bm r',\omega_n)&=&
\left[\left[G_\sigma^0(\bm r,\bm r',\omega_n)\right]^{-1}
-\Sigma_\sigma(\bm r,\bm r',\omega_n)\right]^{-1},
\end{eqnarray}
where $\chi_\mathrm{SC}$ is the pairing susceptibility,
$G_\sigma$ ($G_\sigma^0$) the (non-interacting) Green's function,
$\Sigma_\sigma$ the self-energy, $\omega_n=(2n+1)\pi/T$ the Matsubara frequencies,
and $C(\bm r)$ the space-dependent regularization factor, which is obtained as
\be
C(\bm{r}; \omega_\mathrm{c}; E_\mathrm{Max})
=\pi^{-3}\int_{0}^{\sqrt{2(E_\mathrm{Max} - V(\bm{r}))}} \arctan
\left(\frac{2\omega_\mathrm{c}}{k^2}\right) \mathrm{d}k.
\ee

We discretize the system and use the rotational symmetry of the
system to use a Fourier
component expression in the relative azimuthal angle between
two spatial lattice points.
We need about $15$ (positive) Matsubara frequencies,
$E_\mathrm{Max}\sim30\hbar\overline{\omega}$
and $30-60$ Fourier components for convergence
at $\mu_\uparrow=10\hbar\overline{\omega}\geq\mu_\downarrow$
and $\lambda=1$ or $12$.

If the phase transition from a normal gas to a superfluid is due to the
divergence of the $T$-matrix, the maximum eigenvalue $\Lambda_\mathrm{Max}$
of $\chi_\mathrm{SC}$ reaches zero from below at the transition point $T_\mathrm{c}$.
While in our trapped, finite-size system a first-order transition may occur,
and then $\Lambda_\mathrm{Max}$ is not necessarily zero,
we believe that the transition should happen at
similar values of $\Lambda_\mathrm{Max}$ close to zero
regardless of the trap aspect ratio $\lambda$ if the total number $N$
and temperature $T$ are similar.
Therefore, we compare $\Lambda_\mathrm{Max}$ as a function of $\mu_\downarrow$
for a fixed $\mu_\uparrow$ and $T$.

As shown in Fig. \ref{fig:RSTA}, for $\lambda=12$, the value of
$\Lambda_\mathrm{Max}$ is close to, but does not exceed,
that for $\lambda=1$.
This comparison is for aspect ratios smaller than
those of MIT and Rice; however, because the effects of the trap
shape are enhanced for smaller $N$, this result indicates
that in the equilibrium, for $N\gg10^4$ atoms,
the transition temperatures for a given $P$ for $\lambda=50$
should not exceed that for $\lambda=5$,
and strengthens our conclusion that
$P_\mathrm{CC}$ is not enhanced as $\lambda$ is increased.


To summarize, we have studied superfluidity of population-imbalanced
fermions trapped in an axisymmetric harmonic trap by means of the
Bogoliubov-de Gennes method.
Our numerical results
reproduce the major features of the experiments conducted at MIT,
but does not reproduce those at Rice,
as to the value of $P_\mathrm{CC}$ and as to the degree of LDA
breakdown.
Recently Nascimb\`ene and coworkers at ENS Paris \cite{Nascimbene09}
have trapped population-imbalanced fermions in elongated
traps with various values of the aspect ratio and
observed $P_\mathrm{CC}=0.76(3)$ and no deformation of density profiles.
Zwierlein and coworkers at MIT \cite{Zwierlein09}
observed very long spin diffusion time in two-species fermionic gases at unitarity
by making two polarized gases collide in a
quasi one-dimensional potential.
The diffusion constant extracted from their experiment
suggests that the timescale of the equilibration is as long as one second
for the configuration of the Rice experiment,
which is much longer than the waiting time after the potential ramp in that experiment. 
We speculate that in the Rice experiment for $P>0.8$,
a non-equilibrium condensate state,
possibly from the mechanism discussed in \cite{Parish09},
was observed in the course of slow relaxation and cooling process into a normal state.

M.T. and M.U. gratefully acknowledge fruitful discussions with
R.~Hulet,
W.~Ketterle,
C.~Salomon,
S.~Stringari,
and M.~Zwierlein.
M.T. would like to thank
L.~Baksmaty,
A.~Bulgac,
R.~Ikeda,
C.~Lobo,
T.~Mizushima,
and M.~Urban
for valuable comments.

This research was supported by a Grant-in-Aid for Scientific Research
(Grant No. 22340114)
and the Photon Frontier Network Program of the Ministry of Education, Culture, Sports,
Science and Technology, Japan.
M.T. was supported by Research Fellowship
of the Japan Society for the Promotion of Science for Young Scientists.
Part of the computation in this work has been done using the facilities of
the Supercomputer Center, Institute for Solid State Physics, University of Tokyo.
\end{document}